\documentstyle[aps,epsf]{revtex}

\begin{document}

\title{\large \bf  A new non-Fermi liquid fixed point}
\author{Tae-Suk Kim$^{a,b}$, L. N. Oliveira$^{a,c}$, and D. L. Cox$^{a,d}$}
\address{$^a$Department of Physics, Ohio State University, Columbus, 
Ohio 43210, USA\\
$^b$ PO-Box 118440, Department of Physics, University of Florida,
	 Gainesville, FL, 32611-8440$^*$\\
$^c$ Instituto de F\'{\i}sica de S\~ao Carlos, University of S\~ao
Paulo, C.~Postal 369, 13560-250 S\~ao Carlos, SP, Brazil$^*$\\
$^d$ Institute for Theoretical Physics, University of California,
Santa Barbara, CA 93106-4030\\}
\date{\today}
\maketitle
\begin{abstract}
 We study a new exchange interaction in which the conduction electrons
with pseudo spin $S_c=3/2$ interact with the impurity spin $S_I=1/2$.
Due to the overscreening of the impurity spin by higher conduction electron
spin, a new non-trivial intermediate coupling strength 
fixed point is realized. Using the numerical 
renormalization group (NRG), we show that the low-energy spectra
are described by a non-Fermi liquid excitation spectrum.  
A conformal field theory 
analysis is compared with NRG results and excellent agreement is
obtained.  Using the double fusion rule to generate the operator
spectrum with the conformal theory, we find that the specific heat
coefficient and magnetic susceptibility will diverge as $T^{-2/3}$, 
that the scaling dimension of an applied magnetic field is $5/6$, 
and that exchange anisotropy is always relevant.  
We discuss the possible relevance of our work to two-level system Kondo
materials and dilute cerium alloys, and 
we point out a paradox in understanding the Bethe-Ansatz solutions to
the multichannel Kondo model. 

$^*$ Present address.

\end{abstract}


\section{Introduction}

A number of Ce and U based 
heavy electron metals have recently been discovered to 
show ``non-Fermi liquid physics,'' in such anomalous properties as: 
i) a specific heat coefficient linear in $\log T$, $T$ the 
temperature; ii) resistivity 
which approaches its $T=0$ value as $T^{\alpha}$ with $\alpha$ close
to 1; iii) a magnetic susceptibility $\chi(T)$ which either diverges 
as a logarithm or weak power or saturates as $\chi(T) 
\simeq \chi(0)[1-A\sqrt{T}]$\cite{hfnfl,stegnew,steggoa}. In the case
of UBe$_{13}$ and CeCu$_2$Si$_2$, a superconducting instability arises
within this exotic normal state\cite{stegnew,steggoa}.  
It is an 
open question whether these anomalous properties arise from 
distortions of Fermi liquid physics such as arise in the vicinity of 
a quantum critical point\cite{qcpnfl} or due to a disordered
distribution of Kondo scales\cite{disknfl}, or whether the anomalies
signal the appearance of an entirely new fixed point regulating the 
low temperature physics of these materials.  

In this latter regard, 
it is prudent to study impurity models, such as the overcompensated 
multi-channel 
Kondo model\cite{nozbland} which have very plausible relevance to 
Ce and U based systems\cite{hfnfl,coxold,coxsend}.   In this model, 
$M$ identical spin 1/2 
conduction bands exchange couple to a single spin $S_I$ impurity with the 
condition $M/2 >S_I$, so that there is more conduction spin than needed
to fully compensate the impurity, which leads to an instability of the 
strong coupling (exchange coupling $J\to\infty$) fixed point, as
indicated in Fig. 1.  Because the weak coupling ($J=0$) fixed point is 
unstable always due the Kondo effect, this implies the presence of a 
non-trivial fixed point at intermediate coupling.  This non-trivial 
fixed point corresponds to a degenerate ground state and non-Fermi 
liquid excitation spectrum--no 1:1 map exists of the states to a
non-interacting fermion system.  
In particular, the charge and spin relations relevant for free
fermions do not apply for these systems where spin/charge/channel
separation occurs\cite{al91}.   For more than one impurity,
this fixed point cannot remain stable at low temperatures\cite{2ch2imp},
due to the overlap of the polarization clouds around each site which
have diverging length scales as $T\to 0$.
It is conceivable that the
impurity models may regulate the physics of alloys with finite
concentrations of Ce and U ions or even concentrated compounds 
over a finite range of 
temperatures before ultimately giving way to new fixed points as
$T\to 0$.  This picture is supported by recent results in infinite
spatial dimensions\cite{coxinfd,jarretal}.

Recently, impurity models containing 
a {\it single} high spin electron band have been derived which have been
argued to show non-trivial physics.   These models
have been argued to describe various {\it unstable} fixed points of 
the two-level system Kondo model (describing electron assisted tunneling
of atoms in a double well)\cite{zarand}, and a Ce impurity in a metal
when finite Coulomb interaction is retained\cite{kimdiss,kimcoxscale}.  
The simplest example of this kind of model is for impurity spin $S_I=1/2$ 
and conduction spin $S_c=3/2$.  Although only one channel of conduction
electron is present, this model gives a nontrivial fixed point for the
same reasons as the multi-channel model, as illustrated in Fig. 2.  The
point is that a single large spin $S_c|_{tot}=2$ is coupled to the
impurity as the lowest energy state in strong coupling.  Again, strong 
coupling is unstable, and weak coupling is unstable, so that a
non-trivial intermediate coupling fixed point must exist.  This 
kind of fixed point has been studied with a perturbative scaling 
analysis\cite{zarand,kimdiss,kimcoxscale}, and more recently with 
conformal field theory analysis\cite{senkim}.  

In this paper we present a comprehensive numerical renormalization group
analysis of the $S_I=1/2,S_c=3/2$ single channel Kondo model combined 
with a conformal field theory study of the energy and operator spectrum.
We confirm the existence of a non-trivial fixed point with a non-Fermi 
liquid excitation spectrum.  We find good agreement between finite size
spectra predicted by the conformal theory with those obtained from the 
NRG, both from the group theoretic basis we use for our conformal theory
and that of Ref. \cite{senkim} which uses a different basis.  
Using the double fusion rule of Affleck and Ludwig\cite{al91}, 
we work out the spectrum of primary
field operators for the model, and find that the magnetic susceptibility
and specific heat coefficient in this model will diverge as $T^{-2/3}$
for $T\to 0$, that the scaling dimension of the applied magnetic field
is 5/6, and that exchange anisotropy will always be relevant in this
model.  In concluding, we briefly discuss the possible relevance of this
model to intermediate temperature regimes of two-level system Kondo
materials and dilute cerium alloys.  
Finally, we discuss a separate point of theoretical 
interest concerning a paradox about 
the connection of this model to the Bethe-Ansatz 
solutions of the overcompensated multi-channel Kondo
model\cite{tsvwieg,anddest}.

Our paper is organized as follows:  In Section II we discuss the model
Hamiltonian and the numerical renormalization group methodology used 
to study it.  We also discuss the results from the numerical
renormalization group calculations.  In Section III we develop the
conformal field theory concepts required to study this model and 
calculate the finite size spectra and primary operator spectrum, and we
compare the conformal theory results with the numerical renormalization
group results.
In Section IV we summarize and conclude.

\section{Numerical renormalization group}
 The numerical renormalization group (NRG) technique was invented by
K. G. Wilson to study the Kondo problems. This method defines the 
renormalization group (RG) Hamiltonians and diagonalizes the RG
Hamiltonian numerically. The essence of this method lies in the 
logarithmic discretization of the conduction band and in transforming
the conduction electron Hamiltonian into a tridiagonal form. 
Hence we can visualize the NRG Hamiltonian as a linear array of
sites with the hopping integrals between sites. 
The impurity lies at the $N=-1$ site and the conduction electrons sit at 
the sites $N=0,1,2, \cdots$. 

 Since this method depends on direct numerical diagonalization,
the numerical work requires a huge memory storage compared with,
e.g., the NCA. The consideration of all the 
possible symmetry of the Hamiltonian is essential in reducing 
the computer memory. The NRG is defined by a set of energy eigenvalues
and the reduced matrix elements of the electron operators between 
the energy eigenstates. 

 The NRG approach was the first to show the crossover physics of the 
Kondo effect from the 
high temperature regime to the low temperature fixed point.
K. G. Wilson \cite{wilson}
solved the simple Kondo exchange interaction model,
calculated the magnetic susceptibility, and showed the RG flow
from the high temperature fixed point to the low temperature
fixed point.
Later, the NRG was applied to the simple one-channel Anderson model 
with the conduction electron spins $S_c=1/2$ interacting with 
the impurity system of $S_I=1/2$ \cite{kww}. 
The two-channel Kondo model was also studied using the NRG
 \cite{cragg}. The NRG method was able to show that the two-channel
$S_c=S_I=1/2$ Kondo model leads to the non-Fermi liquid ground
state with a non-trivial finite exchange coupling at the fixed
point. The one-channel two-impurity and two-channel two-impurity 
Kondo problems
have been also studied using the NRG \cite{1ch2imp,2ch2imp}
and the low energy levels from the NRG have been compared
to the finite energy spectrum results of conformal field 
theory to verify the fusion rule hypothesis used 
in the CFT approach \cite{nrganiso}. 

 The NRG is a unique approach for studying stability of various fixed points 
of the Kondo problems. In contrast to other techniques, the NRG
can provide the RG flow diagrams. 
In an earlier work, stability of the two-channel 
non-Fermi liquid fixed points were studied with the NRG 
against the channel anisotropy, the exchange coupling anisotropy, 
and the external magnetic field \cite{nrganiso}. 
This method was extended to the studies of 
the thermodynamics \cite{nrgthermo} and the dynamical properties
 \cite{nrgdynamic1,nrgdynamic2,nrgdynamic3,nrgdynamic4}
like the transport coefficients and the local spectral functions.

\subsection{The conduction electrons with $S_c=3/2$.}
 Here we apply the NRG method to the case that the conduction electrons 
with $S_c=3/2$ interact with the impurity spin $S_I=1/2$.  
Derivation of this exchange interaction is presented in 
Ref\cite{kimcoxscale}. 
Higher conduction electron pseudo spin can overcompensate 
the impurity spin 
leading to the non-Fermi liquid fixed point. In this section, we are going to 
show that this model leads to new non-Fermi liquid fixed point 
using the NRG.

\subsubsection{Logarithmic Discretization.}
 The Model Hamiltonian reads
\begin{eqnarray}
H &=& H_{\rm cb} + H_1, \\
H_{\rm cb} &=& \sum_{\epsilon\alpha} ~\epsilon ~ 
  c_{\epsilon\alpha}^{\dagger}c_{\epsilon\alpha}, \\
H_1 
 &=& J ~ \vec{S}_c (0) \cdot \vec{S}_I. 
\end{eqnarray}
Here the conduction electron spin $S_c=3/2$ and the impurity spin
$S_I=1/2$. We will assume the flat square DOS for the conduction band. 
Following the standard transformation using the logarithmic discretization
 \cite{wilson}, the NRG Hamiltonian reads
\begin{eqnarray}
H_{\rm cb} 
 &=& {1\over 2} ~[~1+\Lambda^{-1}~] ~ 
  \sum_{\alpha} \sum_{n=0}^{\infty} \Lambda^{-n/2} ~
     \xi_n ~[~ f_{n\alpha}^{\dagger} f_{n+1\alpha} + 
        f_{n+1\alpha}^{\dagger} f_{n\alpha} ~], \\
H_{1} &=& 2N(0)J~ \sum_{\alpha} 
   f_{0\alpha}^{\dagger} \vec{L}_{\alpha\beta} f_{0\beta}
     \cdot \vec{S}, \\
\xi_n &=& { 1 - \Lambda^{-n-1} \over \sqrt{(1 - \Lambda^{-2n-1} )
         (1 - \Lambda^{-2n-3} )}}, \\
f_{0\alpha} 
 &=& {1\over \sqrt{2}} ~ \int_{-1}^{1} d\epsilon ~ c_{\epsilon\alpha}.
\end{eqnarray}
Here $\vec{L}$ is the canonical matrix representation of spin $J=3/2$.
This Hamiltonian can be envisaged as a linear array of sites with 
an impurity sitting at the site $n=-1$. Only the electrons at the origin 
($n=0$) interact with the impurity spin. This tight-binding form of Hamiltonian
has a nearest neighbor hopping integral whose magnitude depends on
the distance from the impurity site and is progressively reduced
as the sites move away from the origin. 

 We may define the Hamiltonian up to $N$-th site by $H_N$
\begin{eqnarray}
H_N &=& {1\over 2} ~[~1+\Lambda^{-1}~] ~ \sum_{n=0}^{N-1} \Lambda^{-n/2} ~
     \xi_n ~[~ f_{n\alpha}^{\dagger} f_{n+1\alpha} + 
        f_{n+1\alpha}^{\dagger} f_{n\alpha} ~] + H_0, \\
H_{0} &=& 2N(0)J~ f_{0\alpha}^{\dagger} \vec{L}_{\alpha\beta} f_{0\beta}
     \cdot \vec{S}. 
\end{eqnarray}
We can diagonalize this Hamiltonian step by step. 
The strategy is very simple. When the Hamiltonian upto 
the $N$-th site is assumed diagonalized, we add the $N+1$-th electron site
and diagonalize the extended Hamiltonian again. We continue this process
until the structure of low lying energy levels does not change, or, 
when the stable fixed points are accessed.   
Since the hopping integral 
decreases exponentially with site number $N$, 
rapid saturation is expected for 
a reasonable value of $\Lambda$. To make the lowing lying energy of order 
one, we redefine the Hamiltonian,
\begin{eqnarray}
K_N 
 &\equiv& {2 \over 1+\Lambda^{-1}} ~ \Lambda^{(N-1)/2} ~ H_N \nonumber\\
 &=& \Lambda^{(N-1)/2} ~ \left[ \sum_{n=0}^{N-1} 
   \Lambda^{-n/2} ~ \xi_n ~[~ f_{n\alpha}^{\dagger} f_{n+1\alpha} + 
        f_{n+1\alpha}^{\dagger} f_{n\alpha} ~]
  + 2N(0)J~ f_{0\alpha}^{\dagger} \vec{L}_{\alpha\beta} f_{0\beta}
     \cdot \vec{S} \right], \nonumber\\
 && \\
K_{N+1} &=& \Lambda^{1/2} ~ K_N 
    + \xi_N ~\sum_{\alpha} ~[~ f_{N\alpha}^{\dagger} f_{N+1\alpha} + 
        f_{N+1\alpha}^{\dagger} f_{N\alpha} ~]. 
\end{eqnarray}
The last equation defines the renormalization group of the Kondo 
Hamiltonian.
\begin{eqnarray}
K_0 &=& {2 \Lambda^{-1/2} \over 1 + \Lambda^{-1}} ~ H_0
 = \tilde{g} ~ \sum_{\alpha\beta} f_{0\alpha}^{\dagger} \vec{L}_{\alpha\beta} 
    f_{0\beta} \cdot \vec{S}, \\
\tilde{g}
 &=& {4 \Lambda^{-1/2} \over 1 + \Lambda^{-1}}~JN(0) . 
\end{eqnarray}

\subsubsection{Symmetries.}
Now we show that the axial charge and the total spin operators
commute with the Hamiltonian.  We note that the axial charge may 
be more commonly called the isospin, and that this nomenclature is
peculiar to the NRG literature.  

 The conduction band Hamiltonian is 
invariant under the unitary transformation 
$f_{n\alpha} \to U_{\alpha\beta} f_{n\beta}$. Including the exchange 
interaction, the total spin is conserved
\begin{eqnarray}
\vec{J} &=& \sum_{n} f_{n\alpha}^{\dagger} \vec{L}_{\alpha\beta} f_{n\beta} 
 + \vec{S}.
\end{eqnarray}
It can be easily shown that the total spin of the conduction electrons and 
the impurity spin commutes with the conduction band Hamiltonian
\begin{eqnarray}
 ~[~ H_{\rm cb}, ~\vec{J}~] &=& 0.
\end{eqnarray}
Also this spin operator commutes with the usual Kondo exchange interaction.

 The Hamiltonian is invariant under the unitary transformation:
$F_{n\alpha} \to U F_{n\alpha}$
\begin{eqnarray}
F_{n\alpha} 
  &\equiv & \pmatrix{f_{n\alpha} \cr (-1)^n ~ f_{n\alpha}^{\dagger}}.
\end{eqnarray}
This invariance leads to  
the definition of axial charge symmetry operators \cite{jones}
\begin{eqnarray}
Q^{+} &=& k \sum_{n} (-1)^n ~ \epsilon_{\alpha\beta} 
   f_{n\alpha}^{\dagger}f_{n\beta}^{\dagger}, \\
Q^{-} &=& [Q^{+}]^{\dagger} = k \sum_{n} (-1)^n ~ 
   \epsilon_{\alpha\beta}  ~ f_{n\beta}f_{n\alpha}, \\
Q^{z} &=& {1\over 2} ~ \sum_{n\alpha} 
   \left[ f_{n\alpha}^{\dagger}f_{n\alpha} - {1\over 2} \right]. 
\end{eqnarray}
Here $\epsilon_{\alpha\beta}$ is an antisymmetric tensor 
to be determined. 
These axial charge operators commute with the conduction band Hamiltonian.
$$
~[~ H_{\rm cb}, ~ Q^{i}~] = 0.
$$ 
We will determine the prefactor $k$ and the antisymmetric tensor by 
demanding that these axial charge operators satisfy the angular momentum
commutation relations. First they satisfy 
\begin{eqnarray}
&& [~Q^{z}, Q^{\pm}~] = \pm Q^{\pm}. 
\end{eqnarray}
We demand
\begin{eqnarray}
&& [Q^{+}, Q^{-}] = 2Q^{z}. 
\end{eqnarray}
Then we find the following conditions to be satisfied
\begin{eqnarray}
&& k^2 = {1\over 4}, ~~
 [\epsilon^2]_{\alpha\beta} = - \delta_{\alpha\beta}. 
\end{eqnarray}
Hence we may choose, e.g., for $S_c=3/2$, 
\begin{eqnarray}
k &=& {1\over 2}; ~~\epsilon = \sigma^{x} \otimes i\sigma^{y}. 
\end{eqnarray}
Thus 
\begin{eqnarray}
Q^{+} 
 &=& \sum_{n=0}^{\infty} (-1)^n ~(~f_{n3/2}^{\dagger}f_{n(-3/2)}^{\dagger}
     - f_{n1/2}^{\dagger}f_{n(-1/2)}^{\dagger} ~).
\end{eqnarray}
The above axial charge and angular momentum operators commute with 
each other and with the ordinary exchange interaction. 

 We can also generalize the axial charge operators for the conduction 
electron band with any half-integer spin. 
The angular momentum operators are self-evident.
\begin{eqnarray}
Q^{+} &=& {1\over 2} \sum_{n\alpha} (-1)^n ~ (-1)^{s-\alpha} 
  f_{n\alpha}^{\dagger}f_{n\bar{\alpha}}^{\dagger}, \\
Q^{-} &=& [Q^{+}]^{\dagger}, \\
Q^{z} &=& {1\over 2} ~ \sum_{n\alpha} 
   \left[ f_{n\alpha}^{\dagger}f_{n\alpha} - {1\over 2} \right]
   = {1\over 2} ~\sum_{n} (N_n-s-{1\over 2}). 
\end{eqnarray}
Thus the angular momentum operators and the 
axial charge operators commute with both the conduction band Hamiltonian and 
the ordinary Kondo exchange interaction. They also commute with 
each other, and hence provide good quantum numbers for our numerical 
calculations. 

\subsubsection{Irreducible symmetry tensor operators.}
 In this section, we will construct irreducible tensor operators
at each site which are building blocks for new symmetry 
states in the NRG work. 

The angular momentum operators are
\begin{eqnarray}
S_{N}^{z} 
 &=& \sum_{\alpha} ~\alpha ~ f_{N\alpha}^{\dagger} f_{N\alpha}, \\
S_N^{+} 
 &=& \sqrt{3} ~f_{N3/2}^{\dagger}f_{N1/2} + 2~f_{N1/2}^{\dagger}f_{N-1/2}
   + \sqrt{3} ~ f_{N-1/2}^{\dagger}f_{N-3/2}, \\
S_N^{-} 
 &=& \sqrt{3} ~f_{N1/2}^{\dagger}f_{N3/2} + 2~f_{N-1/2}^{\dagger}f_{N1/2}
   + \sqrt{3} ~ f_{N-3/2}^{\dagger}f_{N-1/2}. 
\end{eqnarray}
And the axial charge operators are
\begin{eqnarray}
Q_{N}^{z} 
 &=& {1\over 2} \sum_{\alpha} ~ f_{N\alpha}^{\dagger} f_{N\alpha} -1, \\
Q_N^{+} 
 &=& (-1)^N~(~f_{N3/2}^{\dagger}f_{N-3/2}^{\dagger} 
            - f_{N1/2}^{\dagger}f_{N-1/2}^{\dagger} ~), \\
Q_N^{-} &=& [~ Q_N^{+} ~]^{\dagger}. 
\end{eqnarray}
The electron operators $f_{N\alpha}^{\dagger}$ and 
the hole operators 
$h_{N\alpha}^{\dagger} = (-1)^{N+3/2+\alpha} ~ f_{N\bar{\alpha}}$ 
are irreducible tensors of rank $3/2$  in the angular momentum sector.
That is,
\begin{eqnarray}
~[~S_N^{z}, ~f_{N\alpha}^{\dagger}~] 
 &=& \alpha ~ f_{N\alpha}^{\dagger}, \\
~[~S_N^{+}, ~ f_{N\alpha}^{\dagger} ~]
 &=& L^{+}_{\alpha+1, \alpha} ~ f_{N\alpha+1}^{\dagger}
  = \sqrt{s(s+1)-\alpha(\alpha+1)}~f_{N\alpha+1}^{\dagger}, \\
~[~S_N^{-}, ~ f_{N\alpha}^{\dagger} ~]
 &=& L^{-}_{\alpha-1, \alpha} ~ f_{N\alpha-1}^{\dagger}
 = \sqrt{s(s+1)-\alpha(\alpha-1)}~f_{N\alpha-1}^{\dagger}, \\
~[~S_N^{z}, ~h_{N\alpha}^{\dagger}~] 
 &=& \alpha ~ h_{N\alpha}^{\dagger}, \\
~[~S_N^{+}, ~ h_{N\alpha}^{\dagger} ~]
 &=& L^{+}_{\alpha+1, \alpha} ~ h_{N\alpha+1}^{\dagger}
 = \sqrt{s(s+1)-\alpha(\alpha+1)}~h_{N\alpha+1}^{\dagger}, \\
~[~S_N^{-}, ~ h_{N\alpha}^{\dagger} ~]
 &=& L^{-}_{\alpha-1, \alpha} ~ h_{N\alpha-1}^{\dagger}
 = \sqrt{s(s+1)-\alpha(\alpha-1)}~h_{N\alpha-1}^{\dagger}. 
\end{eqnarray} 
Here $L^{+}$ and $L^{-}$ are the standard $S=3/2$ angular momentum matrices.
Hence we can apply the Wigner-Eckart theorem when we calculate the matrix 
elements between two different angular momentum operator symmetry states.
In the axial charge case, the electron operators and the hole 
operators satisfy the following commutation relations
\begin{eqnarray}
~[~Q_N^{z}, ~h_{N\alpha}^{\dagger}~] 
 &=& - {1\over 2} ~ h_{N\alpha}^{\dagger}, \\
~[~Q_N^{z}, ~f_{N\alpha}^{\dagger}~] 
 &=& {1\over 2} ~ f_{N\alpha}^{\dagger}, \\
~[~Q_N^{+}, ~-h_{N\alpha}^{\dagger}~] 
 &=& f_{N\alpha}^{\dagger}, \\
~[~Q_N^{+}, ~f_{N\alpha}^{\dagger}~] 
 &=& 0, \\
~[~Q_N^{-}, ~f_{N\alpha}^{\dagger}~] 
 &=& -h_{N\alpha}^{\dagger}, \\
~[~Q_N^{-}, ~h_{N\alpha}^{\dagger}~] 
 &=& 0. 
\end{eqnarray}
Thus we can see that the pair 
$(f_{N\alpha}^{\dagger}, -h_{N\alpha}^{\dagger})$
form a tensor of rank $1/2$ in the axial charge sector. Thus we
can apply the Wigner-Eckart theorem to these conjugate pair operators. 

 We now construct the irreducible symmetry operators, $T_N (QR;JM)$, at the 
site $N$ using the above operators. At each NRG site, the following 
three types of irreducible symmetry tensor operators can be 
developed and they generate 16 new symmetry states:  
\begin{enumerate}
\item ${\bf Q=1, J=0}$: We choose the vacuum state and operate the axial charge 
raising operator to find all states
\begin{eqnarray}
T_N (1\bar{1};0) &\equiv& {\bf 1} \\
T_N (10;0) 
 &=& {1\over \sqrt{2}} ~Q_N^{+}~T_N (1\bar{1};0) \nonumber\\
 &=& {(-1)^{N}\over \sqrt{2}} ~\left( f_{N3/2}^{\dagger}f_{N-3/2}^{\dagger}
   - f_{N1/2}^{\dagger}f_{N-1/2}^{\dagger} \right) \\
T_N (11;0) 
 &=& {1\over \sqrt{2}}~ Q_N^{+}~T_N (10;0)  
 = - f_{N3/2}^{\dagger}f_{N1/2}^{\dagger}
     f_{N-1/2}^{\dagger}f_{N-3/2}^{\dagger}. 
\end{eqnarray}

\item ${\bf Q=1/2, J=3/2}$: First we define the $Q=1/2, R=-1/2$ states and find the 
 $R=1/2$ states operating the axial charge raising operator on them
\begin{eqnarray}
T_N ({1\over 2}(-{1\over 2});{3\over 2} m) 
 &=& f_{Nm}^{\dagger}. 
\end{eqnarray}
We already showed that these operators satisfy the proper commutation relations
with the angular momentum operators.
The $R=1/2$ operators are found by operating the axial charge raising 
operator on the above operators 
\begin{eqnarray}
T_N ({1\over 2}{1\over 2};{3\over 2}m)
 &=& Q_N^{+} ~ T_N ({1\over 2}(-{1\over 2});{3\over 2} m) 
 = h_{Nm}^{\dagger} ~T_N (11;0), \\ 
h_{Nm}^{\dagger}
 &=& (-1)^{N+3/2+m} ~f_{N\bar{m}}. 
\end{eqnarray}
As noted before, the hole operators $h_{Nm}^{\dagger}$ satisfy the tensor 
(of rank $3/2$) commutation relations with the angular momentum operators.

\item ${\bf Q=0, J=2}$: The irreducible operators can be found using the commutation 
 relations
\begin{eqnarray}
T_N (0;22) 
 &\equiv& f_{N3/2}^{\dagger}f_{N1/2}^{\dagger}, \\
T_N (0;21) 
 &=& {1\over 2} ~[~S_N^{-}, ~ T_N (0;22)~] 
 = f_{N3/2}^{\dagger}f_{N-1/2}^{\dagger}, \\
T_N (0;20) 
 &=& {1\over \sqrt{6}} ~[~S_N^{-}, ~ T_N (0;21)~] \nonumber\\
 &=& {1\over \sqrt{2}} ~\left( f_{N3/2}^{\dagger}f_{N-3/2}^{\dagger}
  + f_{N1/2}^{\dagger}f_{N-1/2}^{\dagger} \right), \\
T_N (0;2\bar{1}) 
 &=& {1\over \sqrt{6}} ~[~S_N^{-}, ~ T_N (0;20)~] 
 = f_{N1/2}^{\dagger}f_{N-3/2}^{\dagger}, \\
T_N (0;2\bar{2}) 
 &=& {1\over 2} ~[~S_N^{-}, ~ T_N (0;2\bar{1})~] 
 = f_{N-1/2}^{\dagger}f_{N-3/2}^{\dagger}.
\end{eqnarray}

\end{enumerate}

\subsubsection{Construction of the NRG Hamiltonian and the reduced 
matrix elements of the discretized conduction electron operator} 
 16 new irreducible symmetry electron operators are generated 
at each NRG site. These new operators belong to one of three 
tensor operators
$T_N (QR;JM)$ with rank $Q$ in the axial charge sector and
with rank $J$ in the angular momentum sector
\begin{eqnarray}
&& T_N (00;2M), ~~T_N({1\over 2}R;{3\over 2}M), ~~
T_N (1R;00).
\end{eqnarray}
These irreducible tensor operators are basic building
blocks for the construction of new symmetry sector.

 In constructing the NRG Hamiltonian in each symmetry sector
or in calculating the reduced matrix elements of the discretized
conduction electron operators, we have to calculate the following 
matrix elements
\begin{eqnarray}
&& \left\{ \begin{array}{l}
  T_N^{*} (1R_1;00) \\ \\ T_N^{*} (00;2M_1) \\ \\
  T_N^{*} ({1\over 2}R_1; {3\over 2}M_1)
 \end{array} \right\}
 ~f_{N\alpha}^{\dagger} ~
\left\{ \begin{array}{l}
  T_N (1R_2;00) \\ \\ T_N (00;2M_2) \\ \\ T_N({1\over 2}R_2; {3\over 2}M_2)
 \end{array} \right\}. 
\end{eqnarray}
According to the Wigner-Eckart theorem, the non-vanishing combinations are
\begin{eqnarray}
&& T_N^{*} ({1\over 2}R_1; {3\over 2}M_1) ~ f_{N\alpha}^{\dagger} ~
\left\{ \begin{array}{l}
  T_N (1R_2;00) \\ \\ T_N (00;2M_2)
 \end{array} \right\}, \nonumber\\
&& \left\{ \begin{array}{l}
  T_N^{*} (1R_1;00) \\ \\ T_N^{*} (00;2M_1)
 \end{array} \right\}
 ~f_{N\alpha}^{\dagger} ~ T_N({1\over 2}R_2; {3\over 2}M_2). 
\end{eqnarray}
Guided by the Wigner-Eckart theorem, the explicit calculation gives
\begin{eqnarray}
T_N^{*} ({1\over 2}R_1; {3\over 2}M_1) ~ f_{N\alpha}^{\dagger} ~
   T_N (1R_2; 00) 
 &=& \sqrt{3\over 2}~ <{1\over 2}R_1 | {1\over 2}{1\over 2}; 1R_2> ~
     (M_1, \alpha), \\
T_N^{*} (1R_1; 00) ~ f_{N\alpha}^{\dagger} ~
   T_N ({1\over 2}R_2; {3\over 2}M_2) 
 &=& (-1)^{N+3/2-\alpha} ~ <1R_1 |{1\over 2}{1\over 2}; {1\over 2}R_2> ~
   (M_2, \bar{\alpha}), \nonumber\\
 && \\
T_N^{*} ({1\over 2}R_1; {3\over 2}M_1) ~ f_{N\alpha}^{\dagger} ~
   T_N (00; 2M_2) 
 &=& (-1)^{N+1} ~\sqrt{5\over 2} ~ (R_1, {1\over 2}) ~ 
    <{3\over 2} M_1 | {3\over 2}\alpha; 2M_2>, \nonumber\\
 && \\
T_N^{*} (00; 2M_1)~ f_{N\alpha}^{\dagger} ~ 
   T_N ({1\over 2}R_2; {3\over 2}M_2) 
 &=& \sqrt{2} ~(R_2,-{1\over 2}) ~ <2M_1 | {3\over 2}\alpha; {3\over 2}M_2>. 
\end{eqnarray} 
Here $(x,y)$ is the shorthand notation for the Kronecker delta function
$\delta_{x,y}$. 
Using the above equations and the Wigner-Eckart theorem, we can construct  
the NRG Hamiltonian matrix elements at each symmetry sector which are expressed 
in terms of the reduced electron opreator matrix and the $6-j$
symbols. 

 As mentioned before, the numerical renormalization group is reduced to 
book-keeping of two sets of energy eigenvalues and reduced matrix elements 
in the Kondo problems. The reduced matrix elements are required for 
the construction 
of the NRG Hamiltonian in the next NRG site. Hence it is essential to 
keep track of two data sets.  
In the next section, we are going to present our NRG results, 
%compare with the conformal field theory results, 
and give a physical interpretation.

\subsection{NRG results}
 In this section, we are going to present the results of NRG 
calculation for the exchange interaction of the conduction electrons 
with spin $S_c=3/2$ with the impurity spin $S_I=1/2$. 

 First we show the low energy spectra in the weak coupling limit
with the initial coupling $g=10^{-10}$ (free electron limit).
We used the discretization 
parameter $\Lambda = 5$ and the truncation energy $10$.
In this extremely weak coupling limit, the low lying excitation 
energy spectrum should be described by the free electrons. 
As shown in Fig.~\ref{weakrg}, 
the low lying energy levels are evenly spaced.  

 When we increase the bare exchange coupling, we should be able
to observe the crossover from the weak coupling fixed point 
to a nontrivial non-Fermi liquid fixed point. 
In Fig.~\ref{rg001}, we display the RG flow with initial 
exchange coupling $g=0.01$. We can observe
the clear crossover from the 
weak coupling fixed point to new fixed point. 
The low lying excitation energy spectrum at this new fixed point 
is completely different from that of free electron at the weak 
coupling fixed point. That is, this new fixed point 
is {\it non-Fermi liquid fixed point}. 
>From the Fig.~\ref{rg001}, we can see that degeneracies
at weak coupling fixed point are lifted after crossover
at new non-Fermi liquid fixed point. 
 
 Now we present the three sets of figures with initial 
couplings ranging from weak coupling (see Figs.~\ref{rg01}),
coupling close to the fixed point($g^*=0.2$ form the third-order
scaling analysis, see Figs.~\ref{rg02}), 
and strong coupling(see Figs.~\ref{rg05}). 
In these cases, we used the same model parameters as above. 
The low energy structure is completely different 
from that in the zero coupling limit (free electron). 
That is, this interaction model generates
{\it new non-Fermi liquid fixed point}. 
Three different initial couplings flow to the 
same fixed point as can be seen from the fixed
point low energy spectra.  
The ground state is alternating between $(q=0,j=1/2)$ sector and
$(q=0,j=3/2)$ sector with the conduction electron screening 
shell (NRG site $N$). 
For odd $N$, the $(q=0,j=1/2)$ sector is the 
ground state and vice versa. Here $(q=\mbox{axial charge}, 
j=\mbox{angular momentum})$. 
 
In summary, we have shown using the NRG that 
the one-channel $S_c=3/2, S_I=1/2$
exchange interaction model generates non-trivial fixed point
which is completely different from the Fermi liquid fixed 
point. 

\section{Conformal field theory analysis: low-energy spectrum}
In this section, we carry out a conformal field calculation\cite{ia90,al91}
determining the first few states and 
eigenvalues in the spectrum of the low-energy fixed-point.
In contrast with the NRG procedure, the CFT approach starts out with
the model Hamiltonian expressed in real-space coordinates. 
Following Affleck and Ludwig\cite{al91}, we introduce
left-moving field operators $\psi_\alpha(x)$
($\alpha=\pm3/2, \pm1/2$) and rewrite Eqs.\ (2)~and (3) as
\begin{equation}
\label{e4.1}
H_{cb} = \frac{i v_F}{2\pi}\int_{-\infty}^{\infty} dx ~
\psi_{\alpha}^{\dagger}(x)\partial_x \psi_{\alpha} (x),  
\end{equation}
and
\begin{equation}
\label{e4.2}
H_{1} = v_F \rho J \psi_{\alpha}(0)^{\dagger}\vec{L}_{\alpha\beta}
  \psi_{\beta}(0) \cdot \vec{S}_I,
\end{equation}
where $\rho=1/D$ is the density of states.

\subsection{Free electrons}\label{s1.a}
We must next find the Sugawara forms\cite{go86} equivalent to Eqs.
(\ref{e4.1})~and (\ref{e4.2}). These are 
harmonic oscillator-like Hamiltonians quadratic in certain ``current''
operators of the system 
whose spectrum can be determined by the relevant ladder-operator algebra. 
We follow the logic of Affleck and Ludwig\cite{al91}: 
first, decompose the free
electron Hamiltonian in a Sugawara form which separates out the spin 
currents from other currents (e.g., charge) of the system.  Then couple
the electron spin to the impurity and generate the new spectra.  The 
interaction with the impurity spin is ``absorbed'' into the spin sector
of the Sugawara form Hamiltonian through a simple completing of the
square.  The excitations of the interacting system are found from those
of the non-interacting system through application of the ``fusion
rule''\cite{al91} which, loosely, corresponds to simple addition of 
the impurity spin to the conduction spins subject to the constraints of
the Kac-Moody spin algebra.  

We first consider the case of anti-periodic boundary conditions for the
conduction states which corresponds to the odd-$N$ NRG iterations, and
gives rise to a non-degenerate free electron ground state.  
The Sugawara Hamiltonian is
constructed from the Fourier components of the conserved
currents---axial charge (or isospin) and spin. We therefore consider a segment
of length $2\ell$, with anti-periodic boundary conditions, and define
a Fourier sequence for each component of the axial charge:
\begin{equation}
\label{e4.3}
Q_{n}^{z} =
\sum_{\alpha}\int_{-\ell}^{\ell}\left[\psi^{\dagger}_{\alpha}(x)\psi_{\alpha}(x)-\frac12\right]\exp(-i\pi nx/\ell)\,dx,
\end{equation}
and
\begin{equation}
\label{e4.4}
Q_{n}^{+} =\left(Q_{n}^{-}\right)^{\dagger}=
\int_{-\ell}^{\ell}\left[\psi^{\dagger}_{3/2}(x)\psi^{\dagger}_{-3/2}(x)-
\psi^{\dagger}_{1/2}(x)\psi^{\dagger}_{-1/2}(x)\right]\exp(-i\pi nx/\ell)\,dx.
\end{equation}

Expressed as combinations of the conduction band operators $c_k$,
these components are
\begin{equation}
\label{e4.3a}
Q_{n}^{z} =
\sum_{\epsilon,\alpha}\left[c^{\dagger}_{\epsilon,\alpha}c_{\epsilon+nv_F\pi/2\ell,\alpha}-\frac12\delta_{n,0}\right],
\end{equation}
and
\begin{equation}
\label{e4.4a}
Q_{n}^{+} =\left(Q_{n}^{-}\right)^{\dagger}=
\sum_\epsilon\left(c^{\dagger}_{\epsilon,3/2}c^{\dagger}_{-\epsilon-n\pi
v_F/2\ell,-3/2}-
c^{\dagger}_{\epsilon,1/2}c^{\dagger}_{-\epsilon-n\pi v_F/2\ell,-1/2}\right).
\end{equation}

Likewise, for the spin current, we have
\begin{equation}
\label{e4.5}
\vec J_{n} =
\sum_{\alpha\beta}\int_{-\ell}^{\ell}\psi^{\dagger}_{\alpha}(x)\vec
L_{\alpha\beta}\psi_{\beta}(x)\exp(-i\pi nx/\ell),dx,
\end{equation}
with $\vec J_0$ the total spin operator for the conduction system,
referenced to the filled Fermi sea.

That the Fourier components $\vec Q_n$ ($n=-\infty\ldots\infty$) obey
the level-$k$ Kac-Moody commutation relations\cite{go86} with $k=k_Q=2$
is easily verified by direct substitution of Eqs. (66-69):
\begin{equation}
\label{e4.6}
\left[Q_n^{a}, Q_m^{b}\right] = i\epsilon^{abc}Q^{c}_{n+m} + 
n\delta_{ab}\delta_{n,-m}.
\end{equation}

Similarly, it is easily shown that 
the components $\vec J_n$ obey the level-$k$
Kac-Moody commutation relations with $k=k_J=10$:
\begin{equation}
\label{e4.7}
\left[J_n^{a}, J_m^{b}\right] = i\epsilon^{abc}J^{c}_{n+m} + 
5n\delta_{ab}\delta_{n,-m}.
\end{equation}
Note that although the level of the axial-charge algebra, $k=2$ is
equal to that for the two-channel impurity spin-$1/2$ Kondo
problem\cite{nrganiso}, the level of the spin Kac-Moody algebra, $k=10$
is substantially higher. Thus, although the single-channel conduction 
spin-$3/2$, impurity spin 1/2 model 
and the two-channel impurity spin-$1/2$ model have the same conduction-band
degeneracy, their low-energy spectra are considerably different.
As pointed out by Sengupta and Kim\cite{senkim}, the Kac-Moody algebra
specified by Eq.~(\ref{e4.7}) is an example of the general result for 
conduction spin $S_c$ which gives $k_J = 2S_c(S_c+1)(2S_c+1)/3$.  

The idea is now to write the Hamiltonian as a quadratic form in the 
axial charge (or isospin) currents and the spin currents.  That this is 
possible is suggested by the Harmonic oscillator like structure of the
commutation relation of, e.g., $J_m^a$ with $H_{cb}$.  Specifically, 
\begin{equation}\label{harmosc}
[H_{cb},J_m^a] = -v_F m{\pi \over \ell} J_m^a .  
\end{equation}  
By {\it assuming} such a separation into quadratic forms, directly
computing the commutation and matching the normalization of the 
quadratic forms to Eq.~(\ref{harmosc}), we can construct the Sugawara
hamiltonian.  It turns out that the normalizations depend only on the
levels of the Kac-Moody algebras in the problem.  

The Kac-Moody levels determine the Sugawara form of the conduction
Hamiltonian\cite{go86}:
\begin{equation}
\label{e4.7a}
{H}_{cb} = \frac{v_F\pi}{\ell}
\left[\sum_{n=-\infty}^{\infty}\frac1{12}:\vec J_{-n}\cdot\vec J_{n}:
+\sum_{n=-\infty}^{\infty}\frac1{4}:\vec Q_{-n}\cdot\vec Q_{n}:\right],
\end{equation}
where $:A:$ indicates normal-ordering of the operator $A$. 

We wish to diagonalize this Hamiltonian. To this end, we consider
first the {\em primitive states} (or {\em highest-weight
states}), which are eigenstates devoid of particle-hole
excitations and satisfying the conditions\cite{go86} $q\le k_Q/2\equiv 1$ and
$j \le k_J/2 = 5$. We note that the combination of these constraints
limits the primitive state spin values 
to $j\le 7/2$.  
For $n>0$, it follows from
Eq.~(\ref{e4.3a}) that the $z$ component $Q_{n}^z$ annihilates
particle-hole excitations, and hence $Q_{n>0}^z|\phi_0\rangle= 0$ for
any primitive state $|\phi_0\rangle$. (For $n>0$, by contrast,
$Q_n^z|\phi_0>$ need not vanish, for $Q_{n>0}^z$ {\em creates\/}
particle-hole excitations).  Likewise, $Q_{n>0}^{+}|\phi_0\rangle=
Q_{n>0}^{-}|\phi_0\rangle=0$, since $Q_n^+$ ($Q_n^-$) creates
(annihilates) electron pairs with one particle above the Fermi level
and the other below.  Finally, $\vec J_{n>0}$ annihlates
particle-hole excitations and thus $\vec J_{n>0} |\phi_0\rangle =0$.
It results that, if $|q,j\rangle_0$ is the primitive state with axial
charge $q$ and spin $j$, then on the right-hand side of
Eq.~(\ref{e4.7a}) only the $n=0$ term contributes to its energy
$E_0(q,j)$:
\begin{equation}
\label{e4.7b}
H_{cb}|q,j\rangle_0 =
\frac{v_F\pi}{\ell}\left(
\frac1{12}J_{0}^2
+\frac1{4}Q_{0}^2\right)|q,j\rangle_0.
\end{equation}
Since as Eqs.~(\ref{e4.3a}--\ref{e4.6}) show, $\vec Q_0$ and
$\vec J_0$ are the conduction-band axial charge and spin
operators, respectively, the primitive-state energies are given by
\begin{equation}
\label{e4.7d}
E_0(q,j) =\frac{v_F\pi}{\ell}\left[ \frac{q(q+1)}4 + 
\frac{j(j+1)}{12}\right].
\end{equation}

To determine the quantum numbers $q$ and $j$ associated with each
primitive state, we have to construct it from the single-particle
levels of the conduction-band Hamiltonian.  Particle-hole symmetry
disposes these levels symmetrically with respect to the Fermi energy
$\epsilon_F$. For even number of levels (The number of levels depends
on the number of lattice sites and on the boundary conditions; for
periodic boundary condition, for instance, the number of levels equal
the number of sites.), then, half of them lie above
$\epsilon_F$ and half below, and the ground state of $H_{cb}$ will be
nondegenerate. This spectrum corresponds to an odd numbered iteration of the 
NRG, and to antiperiodic free-fermion boundary conditions.  
For odd number of energy levels, in contrast, one level---which can
accommodate up to four electrons---must lie at the Fermi energy, and
the ground-state degeneracy is $2^4=16$.  This corresponds to the even
numbered NRG iterations and to free electrons with periodic boundary 
conditions. 

For an even number of levels, the nondegenerate ground state has,
therefore, quantum numbers $q=j=0$. The other primitive states are
generated by filling the first level above $\epsilon_F$ with one or
two electrons (or by vacating the first level below $\epsilon_F$),
corresponding to quantum numbers $(q,j)$ equal to $(1/2,3/2)$,
$(1,2)$, $(0,3)$, and another primary state corresponding to three
elementary excitations, which is $(1/2,7/2)$. All other excited states
are not primary: since $q$ increases as the energy levels are
successively filled, the ones not containing elementary particle-hole
excitations must have $q>1$. The five primary multiplets appear in
Table~\ref{t1}.

The 16-fold degenerate ground state comprises three multiplets: (i)
the triplet $(q,j)=(1,0)$, corresponding to a vacant (or fully
occupied) level at $\epsilon_F$, (ii) the octuplet $(q,j) =
(1/2,3/2)$, corresponding to a single electron (or three electrons) at
$\epsilon_F$, and (iii) the quintuplet $(q,j) = (0,2)$, corresponding
to a doubly occupied level at $\epsilon_F$. Along with the states with
two and three elementary excitations, $(q,j)=(1,3)$ and $(1/2,7/2)$,
respectively, these are the only primitive states, since filling any
level above the Fermi level (or vacating any level below $\epsilon_F$)
makes $q>1$. They are listed in Table~\ref{t2}.

The eigenstates $|\phi\rangle$ of $H_{cb}$ containing one or more
particle-hole excitations or with $q>1$ are called {\em descendant
states}; each such state can be obtained from a primitive state by
successive applications of the raising operators $\vec Q_n$ and $\vec
J_n$ ($n=0,-1,-2,\ldots$). Disposed in order of increasing energy, a
primitive state and its descendants form a {\em conformal tower}.

Two eigenstates belonging to a same tower may have
different quantum numbers $q$ or $j$, and two energies on a same tower
may have different degeneracies. The conformal towers are
nonetheless important, for the commutation relations (\ref{e4.6})~and
(\ref{e4.7}) make the energy spacing in each tower
uniform, the splitting $\Delta E= v_F\pi/\ell$ separating any two
successive energies. It follows that, to calculate the low-energy
spectrum of $H_{cb}$, we need only construct the conformal towers 
associated with
the lowest-lying primitive states. Eq.~(\ref{e4.7d}) gives the energy
of the primitive state at the base of a tower, and the energies of the
descendant states are given by
\begin{equation}
\label{e4.7f}
E_m(q,j) = \frac{v_F\pi}{\ell}\left[ \frac{q(q+1)}4 + 
\frac{j(j+1)}{12} + m\right],
\end{equation} 
where $m$ is a positive integer. Notice that the $q$ and $j$ are the
quantum numbers of the primitive states at the base of the tower, in
general different from the quantum numbers of the descendant states.
The eigenvalues of the descendant states listed in Tables
(\ref{t1})~and (\ref{t2}) were computed with Eq.~(\ref{e4.7f}).

\subsection{Interacting system}
We now consider the interaction term $H_1$. Expressed in terms of the 
spin Fourier components $\vec J_n$, the right-hand side of 
Eq.~(\ref{e4.2}) takes a particularly simple form:
\begin{equation}\label{e4.8}
H_1 = 
\frac{v_F\pi}{\ell}\rho J\vec S_I\cdot\sum_{n=-\infty}^{\infty}\vec J_n.
\end{equation}
To take advantage of the similarity between this expression and the one
defining the Sugawara Hamiltonian, Eq.~(\ref{e4.7a}), it is convenient to
introduce the shifted Fourier components
\begin{equation}
\label{e4.9}
\vec {\cal J}_n = \vec J_n + \vec S_I.
\end{equation}
Since the spins $\vec J$ an $\vec S_I$ have the same [SU(2)] symmetry,
these shifted components follow the (level 10) commutation relations
[Eq.~(\ref{e4.7})] obeyed by $\vec J_n$. Moreover, for $\rho J=
1/6$, Eqs. (\ref{e4.7})~and
(\ref{e4.8}) combine into a single expression for the model
Hamiltonian~(1) that is formally equivalent to the conduction
Hamiltonian~(\ref{e4.7}):
\begin{equation}
\label{e4.11}
{H}_{cb}+H_1 = \frac{v_F\pi}{\ell}
\sum_{n=-\infty}^{\infty}\frac1{12}
:\vec {\cal J}_{-n}\cdot\vec {\cal J}_{n}:
+\sum_{n=-\infty}^{\infty}\frac1{4}:\vec Q_{-n}\cdot\vec Q_{n}
\end{equation}

Identifying this special coupling constant with the low-temperature
fixed point of the model Hamiltonian, we reduce the computation of the
low-energy spectrum to the diagonalization of a quadratic Hamiltonian.
This diagonalization follows the steps outlined in Section~\ref{s1.a}.
There, however, the quantum numbers $q$ and $j$ of the conformal towers
were determined by the free-electron single-particle levels. Here, to
determine the quantum numbers $q'$ and $j'$ of the interacting system,
we refer to Affleck's and Ludwig's fusion rule\cite{al91}.

According to that rule, the embedding of the impurity spin $S_I=1/2$
in the conduction-band spin dictated by Eq.~{\ref{e4.7a} leaves
the conformal tower axial charge unchanged, $q'=q$ and yields
a sequence of spins $j'$ ranging from $|j-1/2|$ to the minimum
of $j+1/2$ and $k_J-j-1/2$. Thus, for $j\le9/2$, i.~e., for the spins
we are interested in, the fusion rule is equivalent to the angular
momentum addition law.

It is therefore a simple matter to construct the conformal towers for
the interacting system, the low-energy spectrum being given by
Eq.~(\ref{e4.11}). Results for the cases of a nondegenerate and of a
degenerate free-electron ground state appear in Tables~\ref{t3} and
\ref{t4}, respectively.

Finally, we briefly address an issue raised by Sengupta and
Kim\cite{senkim}, who suggest that for arbitrary $S_I,S_c$, since the 
spin current Kac-Moody algebra is the same as for a multichannel model 
with $k_j = 2S_c(S_c+1)(2S_c+1)/3$ channels that overcompensation will 
occur for impurity spins obeying $S_I<k_j/2$.  This is based upon the
result for the multichannel model which says overcompensation occurs 
for $S_I<k$, $k$ the number of channels.  We disagree with this claim.
To see why, we note that based upon the single fusion rule
the maximal ground state spin which is obtained considering
both non-degenerate and degenerate free spectra is $S_I$(degenerate) 
and $S^*_c-S_I$(non-degenerate)
where
\begin{equation}\label{sstareven}
S^*_c = \sum_{m>0}^{S_c} m = {S_c(S_c+1)\over 2} ~~[2S_c ~even]
\end{equation}
or
\begin{equation}\label{sstarodd}
S^*_c = {(S_c+1/2)^2\over 2} ~~[2S_c ~ odd]~~.
\end{equation}
In view of this, we believe the maximum overcompensated spin value is
determined by the condition $S^*_c>S_I$, not $k_j/2>S_I$.  This checks
with the generalization of the strong coupling picture presented in 
Fig. 2, since the largest
conduction spin you can form from the single band in this picture is in
fact $S^*_c$, which restricts overcompensated impurity spins to 
$S_I\le S^*_c-1/2$.   

\subsection{Boundary Operator Spectra and Physical Properties} 

The conformal theory may also be used to generate the spectra of
boundary 
operators about the non-trivial fixed point by evaluating the spectrum
for a different set of boundary conditions, as noted by 
Affleck and Ludwig\cite{al91}.  Specifically, one considers
the situation where a Kondo impurity is placed at each end of the
system.   Then one applies a {\it double fusion rule} to identify the
spectrum, because the impurity spin must be absorbed at each end.  
Denoting the impurities at either end by $S_I,S_I'$ the double fusion 
rule that yields allowed $S$ values of this double impurity system 
is, for a parent state with spin $S'$ 
\begin{equation}\label{dfus}
0\le S \le min\{S'+S_I+S_I',k_J-S'-S_I-S_I'\} ~~.
\end{equation}

The quantum numbers of the states with these boundary conditions then
identify the quantum numbers of allowed boundary operators.  These 
operators are so called because the critical behavior is limited solely
to the boundary of the system.  The normalized 
energy levels $\ell E/v_F\pi$ give the scaling dimensions $\Delta_o$ of the
operators, which enter into time dependent 
correlation functions of the boundary operators through  the long time
expression
\begin{equation}\label{opcorr}
<O(\tau)O(0)> \sim {1\over \tau^{2\Delta_o}} ~~.
\end{equation}

We present the spectrum of primary field 
boundary operators in Table ~\ref{t5} as determined by the double fusion
rule.  From this table we can glean immediately a number of facts about
the physical properties of the model:\\
1) {\it Low Temperature Behavior of the Specific Heat and Magnetic 
Susceptibility}.  The field with quantum numbers $q=o,j=1$ is a triplet
of primary spin field operators $\Phi^{x,y,z}$, which has scaling index
of $\Delta_{\phi}=1/6$.  In the manner of 
Affleck and Ludwig\cite{al91}, we may identify the leading irrelevant
operator about the fixed point Hamiltonian as $\vec {\cal J}_{-1}\cdot
\Phi$.  If we carry out second order perturbation theory to compute the
impurity contribution to the 
specific heat coefficient $C/T=\gamma$ and magnetic susceptibility $\chi$
precisely 
as did Affleck and Ludwig for the multichannel Kondo model\cite{al91}, 
we will obtain  $\gamma,\chi \sim T^{2\Delta_{\phi}-1} = T^{-2/3}$. This
conclusion was also reached by Sengupta and Kim\cite{senkim}.  \\
2) {\it Crossover in Applied Magnetic Field} Application of a magnetic 
field is a relevant perturbation which, in analogy with the multichannel
model, will drive the system to a Fermi liquid fixed point associated
with a polarized scatterer\cite{nrganiso}.  
The scaling dimension of the magnetic field
$h$ is $1-\Delta_{\phi}=5/6$ and the crossover scale 
$T_h = h^{6/5}/T_K^{1/5}$.  To
see this, consider adding the local Zeeman term 
\begin{equation}\label{zeeman}
S_{Zeeman} = h^z\int_0^{\beta}d\tau \Phi^z(\tau)
\end{equation}
to the action.  Upon rescaling time by $\Lambda$, the prefactor becomes
$h^z\Lambda^{1-\Delta_0}$ due to the singular behavior of $\Phi^z$ for
long times.  This must be a constant to preserve the form of the 
action.  Hence $(h)^{1/(1-\Delta_{\phi}} = (h^z)^{6/5} \sim \Lambda$
and the physical properties will be universal functions in
$x=h^{6/5}/T$. \\
3) {\it Relevance of Exchange Anisotropy} The operator ${\cal Q}$ with
quantum numbers $q=0,j=2$ is a local quadrupolar field.  Because its 
scaling index is 1/2, this suggests that application of a uniaxial
stress or electric field gradient conjugate to the quadrupolar field
will be a relevant perturbation.   This is even the case for
$S_I=1/2,3/2$ in contrast to the multichannel Kondo model.  The reason
is straightforward:  exchange anisotropy in this model will split the
conduction band degeneracy between $|S^z|=3/2,1/2$ states which is 
responsible for the nontrivial ground state.  This is analogous to the 
lifting of channel degeneracy which is always relevant in the
multi-channel Kondo model.   In the event that we generalize from 
impurity spin 1/2 to impurity spin 1, the exchange anisotropy will also
crystal field split the impurity spin and further alter the fixed point
structure.  \\
4) {\it Presence of Fermion Operators}  The operators with $q=1/2,j=3/2$
have appropriate quantum numbers and scaling indices to be identified
with fermion fields.  As in the two-channel model\cite{al91} there are
two such fields here, and the interpretation of this is unclear, though
it appears to be a generic result for overcompensated Kondo models with
$S_I=1/2$.

\subsection{Comparison between NRG and Conformal Theory}

In the last columns of Tables ~\ref{t3} and ~\ref{t4} are NRG
eigenvalues normalized such that the first excited state energy relative
to the ground state is forced to agree with the conformal theory finite
size spectra.  The scaled Energies are then multiplied by a factor of 
24 to multiply out the lowest common denominator of the conformal theory
spectra.  
NRG iterations for odd number of sites correspond to 
even number of levels and so should be compared with the conformal 
theory spectra for non-degenerate free electrons.  NRG iterations with 
even numbers of sites correspond to odd number of levels and thus should
be compared with the conformal theory spectra for degenerate free
electrons.  

We can readily see from the tables that with few exceptions the
agreement is quite satisfactory.  When discrepancies arise, there 
are three sources, identified in Ref. \cite{nrganiso}: (i) the logarithmic
discretization of the NRG conduction band breaks conformal invariance, 
and will lead to detailed deviations between NRG and conformal theory
spectra  as the energy is raised; (ii) the conformal theory spectra 
themselves are generated at the fixed point, and corrections to scaling
can arise as the energy is raised due to the irrelevant operators about
the fixed point (the NRG takes these into account); (iii) the truncation
of states in each $q,j$ sector of the NRG will lead to systematic 
errors in the eigenvalues as one moves the excitation energy towards 
the cutoff value.  

It is satisfactory that in each sector (degenerate vs. non-degenerate 
free states) we not only have good agreement on the energies but also
find all the states generated by the conformal theory up to the energies
considered.  Hence we have little doubt that the non-trivial 
fixed point identified by absorption of the impurity spin in the conformal 
theory is precisely the same as the fixed point identified by the NRG 
analysis.  

\section{Discussion and Conclusion.}

We have analyzed the $S_I=1/2,S_c=3/2$ single band Kondo model with the
Numerical Renormalization Group (NRG) and conformal field theory.  Our work 
confirms that the model displays a non-trivial intermediate coupling 
fixed point which is 
unstable to the relevant perturbations of an applied magnetic field 
and spin exchange anisotropy.  In particular, we have obtained 
detailed numerical
agreement between the finite size spectra generated by the conformal
theory with those of the NRG, both for degenerate and non-degenerate
free electron spectra.   From the conformal theory, we infer that
the impurity contributions to the 
specific heat coefficient and magnetic susceptibility  will diverge 
as $T^{-2/3}$ and that in the presence of a magnetic field the system
will crossover to a Fermi liquid with a polarized scatterer, with 
crossover scale $T_h=h^{6/5}/T_K^{1/5}$ and that thermodynamic
properties will be universal functions of $x=h^{6/5}/T$.  We have 
argued that when the model is generalized to arbitrary $S_I,S_c$ that
non-trivial intermediate coupling fixed points corresponding to
over-compensation will arise for $S_I<S^*_c$ with $S^*_c$ defined by
Eqs. ~\ref{sstareven}~and \ref{sstarodd}.  We have also generated the operator
primitive boundary operator spectrum of the model employing the double
fusion rule.  Our conformal theory results agree with those of Sengupta
and Kim\cite{senkim} for the finite size spectra and low temperature
thermodynamics despite the different choice of basis for the Sugawara 
hamiltonian in our work.  

We wish to close by discussing the experimental relevance  of our work 
and further theoretical issues raised by it. 

Our work may be of relevance at intermediate temperature scales in 
two-level system Kondo impurity materials as an unstable fixed point.
An experimental example of this may have been realized in metallic point
contact devices\cite{ralph}.  
Such fixed points were first studied by Zar\'{a}nd\cite{zarand}.
Similarly, this fixed point may be relevant at intermediate temperature
scales for dilute alloys of Ce$^{3+}$ ions, such as
La$_{1-x}$Ce$_x$Cu$_2$Si$_2$\cite{lacecusi}, such as argued by
Kim\cite{kimdiss} and Kim and Cox\cite{kimcoxscale}.  

On the theoretical front, two key issues are raised.  First, we are
unable to compute the residual entropy and residual resistivity of this
model using the same methods of Affleck and Ludwig\cite{al91} because 
as far as we are aware no generalization of the Kac-Peterson\cite{kacpet} 
and Verlinde\cite{verlinde} formulae 
exists for $SU(2)$ representations other than 
spin 1/2.  The computation of the residual entropy and residual
resistivity rests upon knowledge of the Kac-Peterson\cite{kacpet} 
and Verlinde\cite{verlinde} formulae 
for the spin 1/2 representation of $SU(2)$.  Such a generalization would be 
of theoretical interest at least for working out the general properties
of this model for arbitrary $S_I,S_c$, though it is unlikely to be of 
great experimental relevance.  

Second, and of greater interest is the connection of this model to the
Bethe-Ansatz solutions of the multi-channel Kondo model, which point
was first made by Zar\'{a}nd\cite{zarand} in his $1/k$ expansion
solution
to the $k$-channel two-level system Kondo model.  In the
approach of Tsvelik and Wiegman\cite{tsvwieg} to the $k$-channel
Kondo model, it is shown that a model with a single band of
spin $k/2$ conduction electrons possessing an interaction to the
impurity spin of a certain polynomial $P(x)$ in $x=\vec S_I\cdot\vec
S_c(0)$
is in fact an integrable
model.  It is claimed that this model has the same
Bethe-Ansatz spectrum in the compensated case as the corresponding
$k$-channel impurity Anderson model in the local moment limit where a
Kondo description applies.  It is
then conjectured that in the overcompensated case the single band
spin $k/2$ model has the same spectrum as the $k$-channel Kondo model.
In the case $S_I$=1/2, $P(\vec S_I\cdot\vec S_c)$ reduces to a simple
Heisenberg exchange form. This claim has obvious physical appeal as
illustrated in Fig. 1, since the strong coupling limit of the model
indeed ``fuses'' the conduction electrons into a single large spin
complex with magnitude $k/2$.  However, some doubt about the conjecture
arises when one considers the non-interacting limit.  In that case,
there are 2$k$ free fermion branches to the $k$ channel model, 
whereas the single channel spin $k/2$ band has $k+1$ free fermion 
branches.  Namely, in the noninteracting limit, the number of degrees
of freedom of the two models simply do not match.  

Andrei and Destrei\cite{anddest} solved the Bethe-Ansatz 
{\it directly} on the multichannel model.  
In their approach, a dynamical fusing of
conduction electrons into a spin $k/2$ object describes the ground
state.   This again would appear to nicely match the strong coupling 
picture of Nozi\`{e}res and Blandin\cite{nozbland}. 
The Bethe-Ansatz equations obtained by Andrei and Destrei\cite{anddest} are
identical to those obtained by Tsvelik and Wiegman\cite{tsvwieg}
on the basis of the 
conjecture described in the preceding paragraph.  Indeed, both 
Bethe-Ansatz solutions yield thermodynamics and finite size spectra 
in excellent agreement with NRG and conformal theory results on the $k$
channel models, giving great confidence in the final Bethe-Ansatz 
equations, arrived at through completely different routes.  
An interpretation of the correspondence is that the 
Tsvelik and Wiegman\cite{tsvwieg} conjectured mapping amounts to an
effort to match the dynamical fusion of Andrei and Destrei\cite{anddest}
to a local Hamiltonian. This conjecture leads us to expect that 
the $k=3$ channel $S_I=1/2$ model should have the same spectrum as
the single band $S_c=3/2,S_I=1/2$ model, but this is clearly not the
case as shown here and by Sengupta and Kim\cite{senkim}.  
We regard this discrepancy between
the Tsvelik and Wiegman Bethe-Ansatz approach\cite{tsvwieg} 
and NRG/conformal theory  approaches presented here and in Sengupta
and Kim\cite{senkim} as an interesting paradox to be resolved.

\acknowledgments

 This research was supported by a grant from the U.S. Department of
 Energy, Office of Basic Energy Sciences, Division of Materials
 Research.  The stay of one of us (LNO) at the OSU was financed by a
 fellowship from the CNPq (Brazil).  One of us (DLC) acknowledged the
 support of the Institute for Theoretical Physics, University of
 California Santa Barbara where part of this work was carried out,
 under National Science Foundation Grant PHY94-07194.  We thank
 N. Andrei for many conversations on the comparison between Bethe-Ansatz
 and NRG/conformal theory results, in particular clarifying the
 difference between his\cite{anddest} and the Tsvelik and
 Wiegman\cite{tsvwieg} solutions, as well as pointing out the
 discrepancies between $k$ channel spin 1/2 and spin $k/2$ single
 channel free fermion spectra.  We also thank 
A.M. Tsvelik, J.W. Wilkins, and G. Zar\'{a}nd for
many stimulating interactions.

\appendix

\table
\begin{table}
\caption[ ]{Low-energy spectrum of the free conduction-band Hamiltonian for a
nondegenerate ground state. Energies $E$ (measured from the ground state)
less than $2v_F\pi/\ell$ are shown. The fourth column distinguishes
primary states (p) from descendant ones (d).\label{t1}}
\begin{tabular}{|c|c|c|c|}
$q$ & $j$ & $\frac{\ell E}{v_F\pi}$ & p/d\\\hline\hline
0 & 0 & 0 & p\\\hline
1/2 & 3/2 & 1/2 & p \\\hline
1 & 2 & 1 & p \\
0 & 3 & 1 & p \\
0 & 1 & 1 & d \\
1 & 0 & 1 & d\\\hline
1/2 & 7/2 & 3/2 & p \\
3/2 & 3/2 & 3/2 &  d \\
1/2 & 5/2 & 3/2 & d \\
1/2 & 3/2 & 3/2 & d\\
1/2 & 1/2 & 3/2 & d
\end{tabular}
\end{table}

\begin{table}
\caption[ ]{Low-energy spectrum of the free conduction-band Hamiltonian for a
degenerate ground state. Energies $E$ (measured from the ground state)
less than $v_F\pi/\ell$ are shown. The fourth column distinguishes
primary states (p) from descendant ones (d).\label{t2}}
\begin{tabular}{|c|c|c|c|}
$q$ & $j$ & $\frac{\ell E}{v_F\pi}$ & p/d\\\hline\hline
0   & 2   & 0 & p\\
1/2 & 3/2 & 0 & p \\
1   & 0   & 0 & p \\\hline
1   & 3   & 1 & p \\
1/2 & 7/2 & 1 & p \\
0   & 3   & 1 & d \\
1   & 2   & 1 & d \\
1   & 1   & 1 & d \\
1   & 0   & 1 & d \\
0   & 2   & 1 & d \\
0   & 1   & 1 & d \\
0   & 0   & 1 & d \\
3/2 & 3/2 & 1 & d \\
1/2 & 5/2 & 1 & d \\
1/2 & 3/2 & 1 & d \\
1/2 & 3/2 & 1 & d \\
1/2 & 1/2 & 1 & d
\end{tabular}
\end{table}

\begin{table}
\caption[ ]{Low-energy spectrum of the interacting model Hamiltonian for a
nondegenerate conduction-band ground state (even iteration in NRG). 
Energies of the states
derived from the ones in Table~\ref{t1} by the fusion-rule procedure 
described in the text are shown. The fourth
column distinguishes primary states (p) from descendant ones (d).
The fifth column lists some NRG low energy spectra normalized such that
the energy difference between the ground state and the first 
excited state 
agree with the conformal theory.   Overall the agreement is excellent. 
The lack of perfect numerical agreement is due to the combination of the
breaking of conformal symmetry by the NRG logarithmic discretization
scheme and the truncation of states above a particular energy necessary
to render the NRG block Hamiltonian matrices finite. 
Note the reversal of the pairing of integer(half-integer) $q$ 
to integer(half-integer) $j$ compared to Table \ref{t1}, which indicates
the non-Fermi liquid character of this spectrum in conjunction with the
fractional energy spacings.\label{t3}}
\begin{tabular}{|c|c|c|c|c|}
$q$ & $j$ & $\frac{\ell E}{v_F\pi}$ & p/d & $E_{NRG}(/24)$ \\ \hline \hline
0 & 1/2 & 0 & p & 0 \\\hline 
1/2 & 1 & 7/24 & p & 7.000 \\\hline 
1/2 & 2 & 5/8 & p  & 15.166 \\\hline 
0 & 5/2 & 2/3 & p & 16.018 \\\hline 
1 & 3/2 & 3/4 & p  & 18.051 \\\hline
0 & 3/2 & 1 & d & 24.200 \\ 
0 & 1/2 & 1 & d & 24.358\\
1 & 1/2 & 1 & d & 24.262\\\hline
1 & 5/2 & 7/6 & p  & 28.443 \\\hline
0 & 7/2 & 5/4 & p & 30.309\\\hline
3/2 & 1 & 31/24 &  d & 31.165\\
1/2 & 2 & 31/24 & d & 31.486\\
1/2 & 1 & 31/24 & d & 31.198\\\hline
3/2 & 2 & 13/8 &  d & 39.637\\
1/2 & 1 & 13/8 & d & 35.077\\
1/2 & 0 & 13/8 & d & 31.133\\\hline
0 & 7/2 & 5/3 & d & 40.443\\
0 & 5/2 & 5/3  & d & 45.446 \\
0 & 3/2 & 5/3 & d & 40.581\\
1 & 5/2 & 5/3 & d & 41.154
\end{tabular}
\end{table}

\begin{table}
\caption[ ]{Low-energy spectrum of the model Hamiltonian for a
degenerate conduction-band ground state (odd iteration in NRG). 
Energies of the states
derived from the ones in Table~\ref{t2} by the fusion-rule procedure 
described in the text are shown. The fourth
column distinguishes primary states (p) from descendant ones (d).
The fifth column lists some NRG low energy spectra normalized such that
the energy difference between the ground state and the first 
excited state 
agree with the conformal theory.  Overall the agreement is
excellent. The lack of precise numerical agreement is due
to a combination of the breaking of conformal symmetry by the
NRG logarithmic discretization procedure and the truncation of 
energy levels to render the NRG block Hamiltonian 
matrices finite for diagonalization.
Note the reversal of the binding of integer(half-integer)$q$ to 
integer(half-integer)$j$ in comparison with Table \ref{t2}, which, along
with the fractional energy spacings, indicates the inapplicability of 
Fermi liquid theory to this spectrum.\label{t4}}
\begin{tabular}{|c|c|c|c|c|}
$q$ & $j$ & $\frac{\ell E}{v_F\pi}$ & p/d & $E_{NRG}(/24)$\\ \hline\hline
0   & 3/2 & 0    & p & 0.000\\\hline 
1/2 & 1   & 1/24 & p & 1.000 \\\hline 
1   & 1/2 & 1/4  & p & 6.004 \\\hline 
1/2 & 2   & 3/8  & p & 9.014 \\\hline 
0   & 5/2 & 5/12 & p & 10.018 \\\hline
1/2 & 3   & 7/8  & p & 21.782 \\\hline 
1   & 5/2 & 11/12 & p & 22.759 \\\hline
0   & 5/2 & 1   & d &  24.715\\
1   & 3/2 & 1    & d &  24.942\\
0   & 3/2 & 1   & d & 25.245 \\
0   & 1/2 & 1    & d & 24.879 \\\hline
3/2 & 1   & 25/24& d & 25.593 \\ 
1/2 & 2   & 25/24& d & 26.114 \\
1/2 & 1   & 25/24& d & 25.550 \\
1/2 & 1   & 25/24& d & 26.608\\\hline
1/2 & 0   & 25/24& d & 26.938\\\hline
1   & 1/2 & 5/4 & d & 30.765 \\
1   & 1/2 & 5/4 & d & 33.357 \\	
1   & 3/2 & 5/4 & d & 31.789 \\
0   & 1/2 & 5/4  & d & 31.967 \\\hline
3/2 & 2   & 11/8 & d & 35.504 \\
1/2 & 3   & 11/8 & d & 34.977 \\
1/2 & 2   & 11/8 & d & 34.969 \\
1/2 & 2   & 11/8 & d & 36.339 \\
1/2 & 1   & 11/8 & d & 34.889 \\\hline
1   & 5/2 & 17/12& d & 36.459 \\
0   & 7/2 & 17/12 & d & 35.918\\
0   & 5/2 & 17/12& d & 37.471 \\
0   & 3/2 & 17/12& d & 35.823 \\\hline
\end{tabular}
\end{table}

\begin{table}
\caption[ ]{Primary Field Boundary
Operator Spectrum for the $S_c=3/2,S_I=1/2$ model.  We
obtain this spectrum by applying the double fusion rule of Affleck and
Ludwig\cite{al91} to the spectrum of Table \ref{t3}.  This rule assumes an 
impurity at each end of the finite size chain of length $\ell$.  
The scaling dimensions $\Delta_o$  
are simply the scaled energies $\ell E/v_F\pi$ for this double
impurity boundary condition.  The $Q=0,j=0$
operator is simply the local charge of the site. The physical
interpretation of the $q=0,j=1$ operator is the primary spin field at 
the impurity site. With the scaling dimension of 1/6, this indicates 
that the specific heat coefficient and susceptibility will diverge as 
$T^{-2/3}$ as argued in the text.  
The $q=1/2,j=3/2$ field is a fermionic operator, which
always must have scaling index 1/2 in any model.  Fields with $q=1$ 
include pair fields ($q_z=\pm 1$), which clearly do not provide singular
pair field susceptibilities here. 
The $q=0,j=2$ field is a quadrupolar tensor, which 
given scaling index 1/2, indicates the relevance of exchange anisotropy
for this model.  See the text for further discussion.\label{t5}}
\begin{tabular}{|c|c|c|}
$q$ & $j$ & $\Delta_o$ \\\hline\hline
0   & 0   & 0 \\
0 & 1/2 & 1/6 \\
1/2   & 1/2   & 1/4 \\
1/2  & 3/2 & 1/2 \\
1/2  & 3/2 & 1/2 \\
0  & 2 & 1/2 \\
1  &  1 &  2/3 \\
1/2 & 5/2 & 11/12 \\
1 & 2 & 1 \\
0 & 3 & 1 \\
0 & 3 & 1 \\
1 & 2 & 1 \\
1 & 3 & 3/2 \\
0 & 4 & 5/3
\end{tabular}
\end{table}

\figure
\begin{figure}
%\protect\centerline{\epsfxsize=4.0in \epsfbox{figs/mulcartoon.ps}}
\vskip 1.0cm
\protect\caption[Cartoon of two-channel model at strong
coupling.]
{{\bf Strong coupling limit of the two-channel $S_I=1/2$ Kondo model}
At strong coupling (zero kinetic energy) in the two-channel spin
1/2 Kondo model, two units of conduction spin (labeled $c1$ and $c2$) 
bind to form a net
spin 1 complex that lines up antiparallel to the impurity spin.
The resulting spin of the ground state is $S_{tot}=1/2$.  The
strong coupling fixed point is unstable to the introduction of 
the kinetic energy because antiferromagnetic 
superexchange will be generated to electrons off the impurity
site as argued by Nozi\`{e}res and Blandin\cite{nozbland} of order 
$t^2/J$, where $t$ is the intersite hopping, which 
maps the effective model back to the weak coupling limit where 
the exchange interaction must grow.  Since both weak coupling and strong
coupling limits are unstable, a non-trivial fixed point must exist 
at intermediate coupling.}
\protect\label{multicartoon}
\end{figure}

\begin{figure}
%\protect\centerline{\epsfxsize=4.0in \epsfbox{figs/singcartoon.ps}}
\vskip 1.0cm
\protect\caption[Cartoon of the Single channel $S_c$=3/2, $S_I=1/2$ 
model.]  {{\bf Strong coupling limit of the single channel
$S_c=3/2,S_I=1/2$ model} In the strong coupling limit of this model,
two conduction electrons with $S_z=3/2,1/2$ will bind to form a spin 2
object that aligns antiparallel with the impurity spin.  This strong
coupling fixed point is unstable to the introduction of the kinetic
energy which will generate an antiferromagnetic superexchange with the
$S_{tot}=3/2$ bound object of order $t^2/J$, which then maps it back
to a weak coupling Kondo problem with growing exchange interaction.
Since both strong- and weak-coupling limits are unstable, a
non-trivial fixed point must exist at intermediate coupling.}
\protect\label{singlecartoon}
\end{figure}

\begin{figure}
%\protect\centerline{\epsfxsize=4.0in \epsfbox{figs/gzero.ps}}
\vskip 1.0cm
\protect\caption[RG flow diagram in the free electron limit
for $S_c=3/2, S_I=1/2$ exchange interaction model]
{{\bf RG flow diagram in the free electron limit
for $S_c=3/2, S_I=1/2$ exchange interaction model.}
The RG flow diagram is displayed for bare coupling $g=10^{-10}$.
This bare coupling corrrespond to the free electron limit.
The low lying exciattion energy spectrum is that of free
conduction electrons. Note that low lying energy levels are
evenly spaced. Labeling the first five lowest lying symmetry sectors
are as follows throughout all the RG flow diagrams presented here.
(1) $N=$ odd case. the ground energy symmetry sector is
 $(0,1/2)$: the solid line;
 $(1/2,1)$: the dashed line;
 $(1/2,2)$: the dash-dotted line;
 $(0,5/2)$: the dash-dot-dotted line;
 $(1,3/2)$: the dash-dash-dotted line.
(2) $N=$ even case. the ground energy symmetry sector is
 $(0,3/2)$: the solid line;
 $(1/2,1)$: the dashed line;
 $(1,1/2)$: the dash-dotted line;
 $(1/2,2)$: the dash-dot-dotted line;
 $(0,5/2)$: the dash-dash-dotted line.}
\protect\label{weakrg}
\end{figure}

\begin{figure}
%\protect\centerline{\epsfxsize=4.0in \epsfbox{figs/g001.ps}}
\vskip 1.0cm
\protect\caption[RG flow diagram displaying the crossover from the
weak coupling to non-trivial finite coupling fixed point
for $S_c=3/2, S_I=1/2$ exchange interaction model]
{{\bf RG flow diagram displaying the crossover from the
weak coupling to non-trivial finite coupling fixed point
for $S_c=3/2, S_I=1/2$ exchange interaction model.}
The RG flow diagram is displayed for initial exchange
coupling $g=0.01$. With this weak coupling, we can observe
the clear crossover from the weak coupling fixed point
to {\it new fixed point} which is completely different
from the former.
Before the crossover, the low lying exciattion energy spectrum
is close to that of free conduction electrons.
After the crossover, the low energy excitation spectrum
is described by non-Fermi liquid. See Figs. \ref{rg01}, \ref{rg02},
and \ref{rg05}.}
\protect\label{rg001}
\end{figure}

\begin{figure}
%\protect\centerline{\epsfxsize=4.0in \epsfbox{figs/g01.ps}}
\vskip 1.0cm
\protect\caption[RG flow diagram with the initial weak coupling
for $S_c=3/2, S_I=1/2$ exchange interaction model]
{{\bf RG flow diagram with the initial weak coupling
for $S_c=3/2, S_I=1/2$ exchange interaction model.}
The RG flow diagram is displayed for bare coupling $g=0.1$.
This bare coupling corresponds to the weak coupling regime.
The low lying excitation spectrum is quite different from those
of free electrons in Fig.~\ref{weakrg}.}
\protect\label{rg01}
\end{figure}

\begin{figure}
%\protect\centerline{\epsfxsize=4.0in \epsfbox{figs/g02.ps}}
\vskip 1.0cm
\protect\caption[RG flow diagram with the initial coupling
close to the fixed point
for $S_c=3/2, S_I=1/2$ exchange interaction model]
{{\bf RG flow diagram with with the initial coupling
close to the fixed point
for $S_c=3/2, S_I=1/2$ exchange interaction model.}
The RG flow diagram is displayed for bare coupling $g=0.2$.
This bare coupling is close to the fixed point. Note that
the third-order scaling analysis gives $g^*=0.2$
at the fixed point.}
\protect\label{rg02}
\end{figure}

\begin{figure}
%\protect\centerline{\epsfxsize=4.0in \epsfbox{figs/g05.ps}}
\vskip 1.0cm
\protect\caption[RG flow diagram with the initial strong coupling
for $S_c=3/2, S_I=1/2$ exchange interaction model]
{{\bf RG flow diagram with the initial strong coupling
for $S_c=3/2, S_I=1/2$ exchange interaction model.}
The RG flow diagram is displayed for bare coupling $g=0.5$.
This bare coupling corresponds to the strong coupling regime.}
\protect\label{rg05}
\end{figure}

\end{document}